\documentclass[aps,twocolumn,longbibliography]{revtex4}

\usepackage[dvips]{graphics,graphicx}
\usepackage[colorlinks, linkcolor=blue, citecolor=blue, urlcolor=blue, breaklinks=true]{hyperref}
\usepackage{amsmath}
\usepackage{bm}
\usepackage{color}

\begin{document}
\title{Self-Ordering, Cooling, and Lasing in an Ensemble of Clock Atoms}
\author{Anna Bychek$^{1}$}
\author{Laurin Ostermann$^{1}$}
\author{Helmut Ritsch$^{1}$}
\affiliation{
$^1$Institut f\"ur Theoretische Physik, Universit\"at Innsbruck, Technikerstr. 21a, A-6020 Innsbruck, Austria}
\date{\today}
\begin{abstract}
Active atomic clocks are predicted to provide far better short-term stability and robustness against thermal fluctuations than typical feedback-based optical atomic clocks. However, continuous laser operation using an ensemble of clock atoms still remains an experimentally challenging task. We study spatial self-organization in a transversely driven ensemble of clock atoms inside an optical resonator and coherent light emission from the cavity. We focus on the spectral properties of the emitted light in the narrow atomic linewidth regime, where the phase coherence providing frequency stability is stored in the atomic dipoles rather than the cavity field. The atoms are off-resonantly driven by a standing-wave coherent laser transversely to the cavity axis allowing for atomic motion along the cavity axis as well as along the pump. In order to treat larger atom numbers we employ a second-order cumulant expansion which allows us to calculate the spectrum of the cavity light field. We identify the self-organization threshold where the atoms align themselves in a checkerboard pattern, thus maximizing light scattering into the cavity, which simultaneously induces cooling. For a larger driving intensity, more atoms are transferred to the excited state, reducing cooling but increasing light emission from the excited atoms. This can be enhanced via a second cavity mode at the atomic frequency spatially shifted by a quarter wavelength. For large enough atom numbers we observe laser-like emission close to the bare atomic transition frequency.
\end{abstract}
\maketitle

\section{Introduction}
Building a continuously operating superradiant laser, which constitutes the core of a next-generation active optical atomic clock, has been an integral part of developing more robust and accurate timekeeping devices in the last two decades, both theoretically and experimentally~\cite{Holland09,Thompson12,Maier14,Norcia18,debnath2018lasing,Liu20,Bychek2021,Wu22,Zhang22,Kazakov22,zhang2023extremely,Bohr24,yu2024prospects, yu2024conditional}. The continuous clock operation on a narrow-linewidth atomic transition is of  particular interest for frequency metrology, precision measurements and quantum sensing~\cite{bothwell2022resolving, ye2024essay}.

Various advances in this direction, such as pulsed superradiance~\cite{norcia2016superradiance, laske2019pulse, tang2021cavity}, continuous lasing proof of principle experiments on kHz transitions~\cite{Thompson16, tang2022prospects, kristensen2023subnatural, fama2024continuous}, conveyor belt setups~\cite{Schreck19, escudero2021steady, chen2022continuous, schafer2024continuous, takeuchi2023continuous} or more complicated multilevel pumping protocols~\cite{hotter2022continuous} have been reported in the last years. However, a continuous superradiant laser on an optical transition remains challenging and has not been realized yet.

In contrast to the conventional mechanism of self-organization operating in the far-detuned dispersive regime~\cite{domokos2002collective, chan2003observation, asboth2005self, niedenzu2011kinetic, arnold2012self, ritsch2013cold, schutz2015thermodynamics, jager2016mean, nairn2024spin}, where the atoms are never excited by the pump laser but act as light scatterers into the cavity creating their own trapping potential, we excite the atoms and leverage state-dependent light forces created by the pump laser and the light scattered into the cavity mode. In this way the atoms will acquire an excited state population while subsequently moving through field minima resulting in an almost unperturbed atomic transition frequency.  At this point, the atoms will serve as the gain medium for our laser by means of stimulated emission.

Methodologically expanding our previous work in ref.~\cite{Bychek23}, we employ a second-order cumulant expansion approach~\cite{kubo62, Holland09, plankensteiner2022quantumcumulants} that allows for the treatment of experimentally realistic system sizes, while still capturing the essence of the involved physics as suggested by comparison to full quantum simulations for a small atom number.~In the present work, we focus on the self-ordering of atoms under the transverse coherent drive and the spectral properties of the emitted light in a superradiant regime.~The spatial degrees of freedom are treated in a semi-classical phase space approach where we include atomic motion along the cavity axis as well as along the direction of the orthogonal coherent pump.

\section{Model}
%
%
\begin{figure}[t]
    \centering
	\includegraphics[width=\columnwidth]{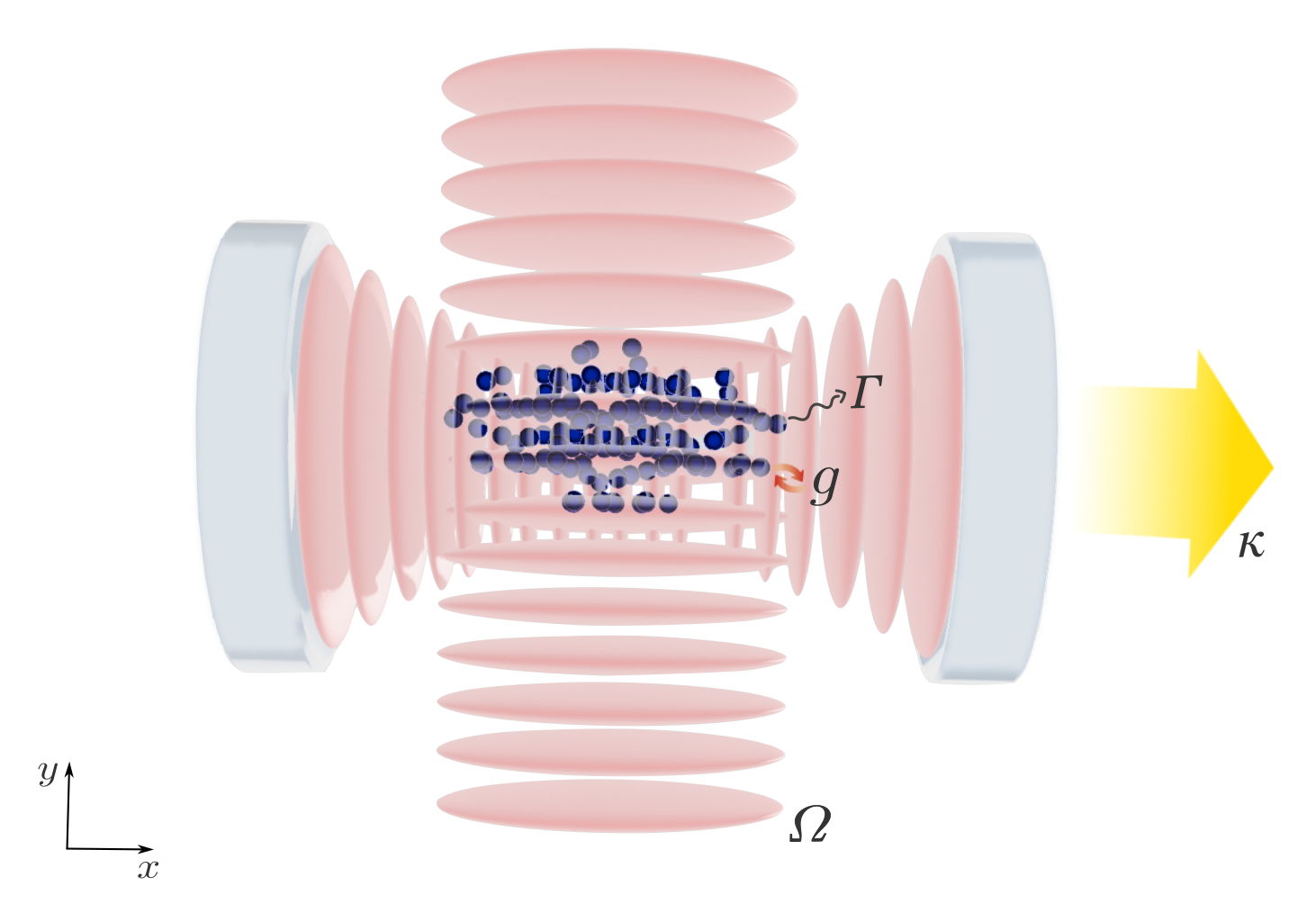}
	\caption{\textit{Schematic illustration of the system.} An ensemble of clock atoms is trapped within a red-detuned standing wave transverse to a Fabry-Perot optical cavity with one mode red-detuned close to the pump field. The atomic transition couples to the cavity ($g$), the pump ($\Omega$) and the environment ($\Gamma$) allowing for collective resonant scattering from the pump to the cavity mode with losses characterized by a decay rate ($\kappa$). Light forces via collective scattering will modify atomic motion in the $xy$-plane and eventually induce 2D atomic ordering, cooling and lasing.}
	\label{fig0}
\end{figure}
%
%
We study the dynamics and self-organization of atoms having a narrow atomic linewidth in a so-called intermediate bad-cavity regime, where the cavity linewidth ($\kappa$) is bigger than the linewidth of the gain medium ($\Gamma$). In the limit of small photon numbers, the phase coherence accounting  for frequency stability is stored in the atoms rather than in the cavity field. In the absence of an incoherent pumping mechanism for the atoms, we aim for conditions under which the atomic dynamics in the cavity leads to continuous lasing close to the bare atomic transition frequency.

%
\begin{figure*}[t]
    \centering
	\includegraphics[width=0.8\textwidth]{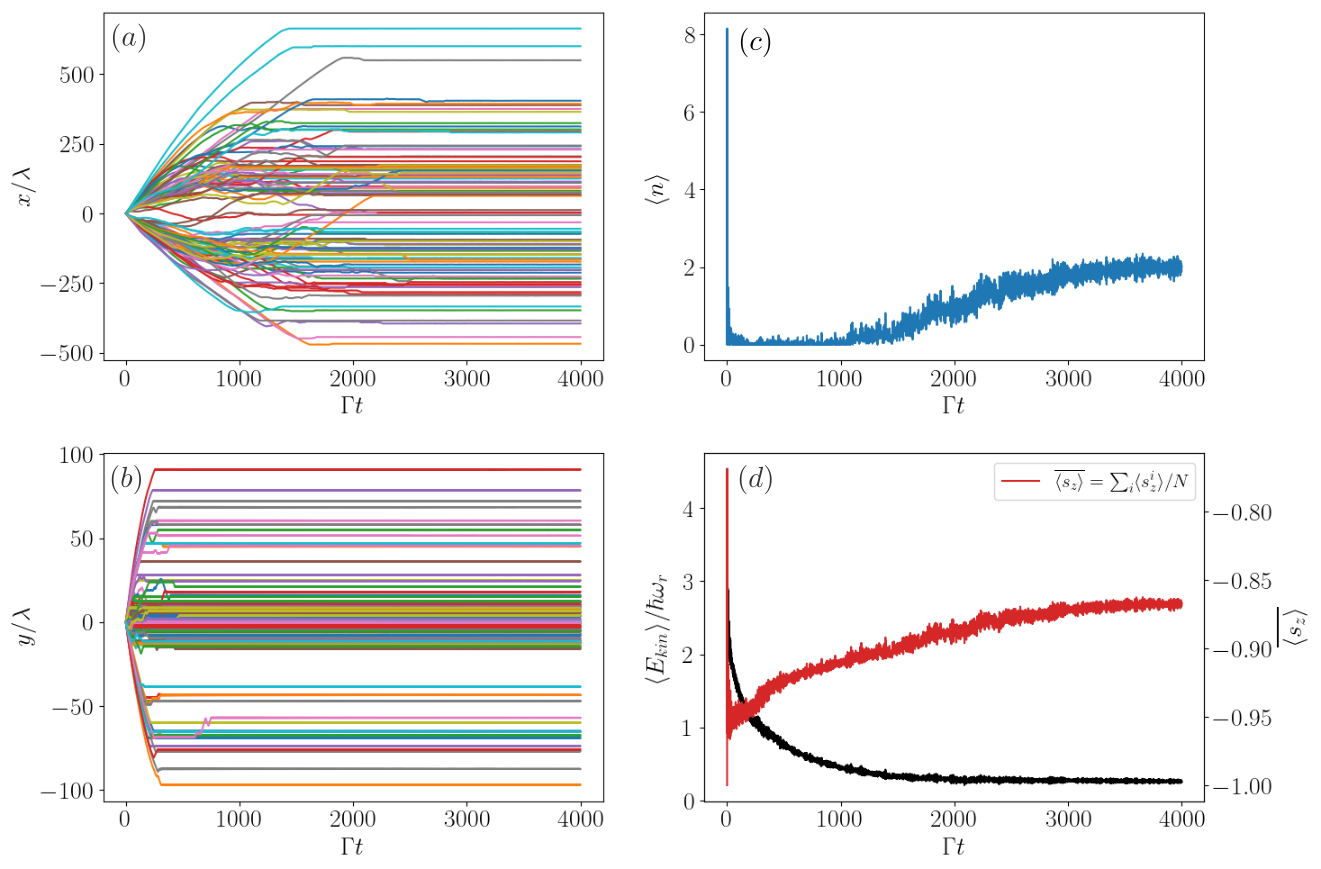}
	\caption{\textit{Dynamics.} (a-b) (x,y) motion of $N=100$ atoms in the parameter regime of self-organization, (c) mean intra-cavity photon number in the course of time, (d) mean kinetic energy (black line) and inversion averaged over the atomic ensemble (red line) depending on time. Parameters: $g=1\Gamma$, $\kappa = 10\Gamma$, $\Omega = 5\Gamma$, $\Delta_a = -20\Gamma$, $\Delta_c = -10\Gamma$, $w_0 = 1000 \lambda$, $\omega_r = 1\Gamma$. The curves show concurrent trapping, cooling and self-ordering via collective light scattering into the cavity mode.}
	\label{fig1}
\end{figure*}
%

We consider a basic model for self-organization in a linear single mode cavity ($\omega_c$), as depicted in fig~\ref{fig0}. The two-level atoms ($\omega_a$) are illuminated off-resonantly ($\omega_{\Omega}$) by a coherent standing-wave laser drive $\Omega \cos{(ky)}$ transversely to the cavity axis and experience a dipole light force affecting their spatial degrees of freedom in the cavity. The phase of the scattered light is determined by the particles position and as the driving strength surpasses the self-organization threshold one can observe the atoms arrange themselves into one of two regular checkerboard patterns, thus maximizing scattering into the cavity. This can be shown from the solution of the master equation for the density matrix ($\hbar = 1$),
\begin{equation}
    \label{master equation}
\dot \rho =  -i \left[ H, \rho\right] + \mathcal{L}_\kappa \left[ \rho \right] + \mathcal{L}_\Gamma \left[ \rho \right],
\end{equation}
with the Hamiltonian in the rotating frame of the laser drive, which can be written as
\begin{equation}
\begin{aligned}
    \label{eq:H}
 H &= -\Delta_a \sum_{i=1}^N \sigma^+_i \sigma^-_i -\Delta_c a^\dagger a  +  \sum_{i=1}^N g(x_i,y_i) (a \sigma^+_i + a^\dagger \sigma^-_i)\\
 &+ \sum_{i=1}^N \Omega \cos(k' y_i) \left( \sigma^+_i + \sigma^-_i \right),
\end{aligned}
\end{equation}
where $\Delta_a = \omega_{\Omega}-\omega_a$, $\Delta_c = \omega_{\Omega}-\omega_c $, $\sigma^+_i = (\sigma^-_i)^\dagger = | e \rangle_i \langle g |_i$ denote the raising (lowering) operators of the $i$-th atom with ground state $| g \rangle$ and excited state $| e \rangle $. The ladder operator $a^\dagger$ ($a$) is the photon creation (annihilation) operator of the cavity mode and atoms experience a position-dependent coupling to the cavity $g(x,y) = g \cos{(kx)} e^{-y^2/w_0^2}$. We describe the atomic motion in the $xy$-plane by semi-classical equations,
\begin{equation}
\begin{aligned}
    \label{eq:motion}
\dot{\langle x \rangle}  = 2\omega_r \langle p_x \rangle / k_a^2, \quad &\dot{\langle y \rangle}  = 2\omega_r \langle p_y \rangle / k_a^2,\\
\dot{\langle p_x \rangle} = -\Big\langle \frac{\partial H}{\partial x}\Big\rangle, \quad &\dot{\langle p_y \rangle} = -\Big\langle \frac{\partial H}{\partial y} \Big\rangle,
\end{aligned}
\end{equation}
where  ${\omega_r = k_a^2/(2m)}$ is the atomic recoil frequency given by the atomic mass $m$ and wave number $k_a = \omega_a/c$ of the atomic transition. Since the laser is not very far detuned from both the cavity and the atomic transition frequency, we suppose that $k \approx k' \approx k_a$ and measure the distances in units of the atomic transition wavelength $\lambda$.

The Liouvillian in the standard Lindblad form represents the decay processes
\begin{equation}
\begin{aligned}
\label{eq:L}
\mathcal{L}_\kappa[\rho] &= \frac{\kappa}{2}(2a\rho a^\dagger - a^\dagger a \rho - \rho a^\dagger a)\\
\mathcal{L}_\Gamma[\rho] &= \frac{\Gamma}{2} \sum_{i=1}^N (2\sigma^-_i \rho \sigma^+_i -\sigma^+_i \sigma^-_i \rho -\rho \sigma^+_i \sigma^-_i)
\end{aligned}
\end{equation} 
describing individual spontaneous atomic emission with the rate $\Gamma$ and cavity losses with the rate $\kappa$.

In the self-organization regime, atoms scatter light into the cavity in such a way that the resulting field simultaneously traps and cools the atoms keeping them in an organized pattern, which is stable on a long time scale. More precisely, in the dispersive regime the following two conditions have to be met~\cite{asboth2005self},
\begin{equation}
\label{condition_1}
\delta = \Delta_c -\frac{1}{\Delta_a}\sum_{i=1}^N g^2(x_i, y_i)<0,
\end{equation} 
stemming from the effective cavity frequency due to the presence of the atoms. Cavity cooling is ensured when the effective cavity detuning $\delta$ is negative. In the case of positive detuning, the atoms acquire kinetic energy from the driving light resulting in cavity heating. The second condition~\cite{niedenzu2011kinetic},
\begin{equation}
\label{condition_2}
\frac{2\sqrt{N} g' \Omega}{|\Delta_a|} > \frac{(\kappa/2)^2 +\delta^2}{2|\delta|},
\end{equation} 
with $g' = \sqrt{\sum_i g^2(x_i,y_i)/N}$ the averaged atom-field coupling over the ensemble, corresponds to the pumping threshold above which the atoms tend to self-organize in order to minimize their energy in the potential resulting from the interference between the cavity and the pump field. In this work, we will examine these conditions as we move away from the dispersive regime towards the saturation of the atomic transition.

\begin{figure}[t]
    \centering
	\includegraphics[width=0.97\columnwidth]{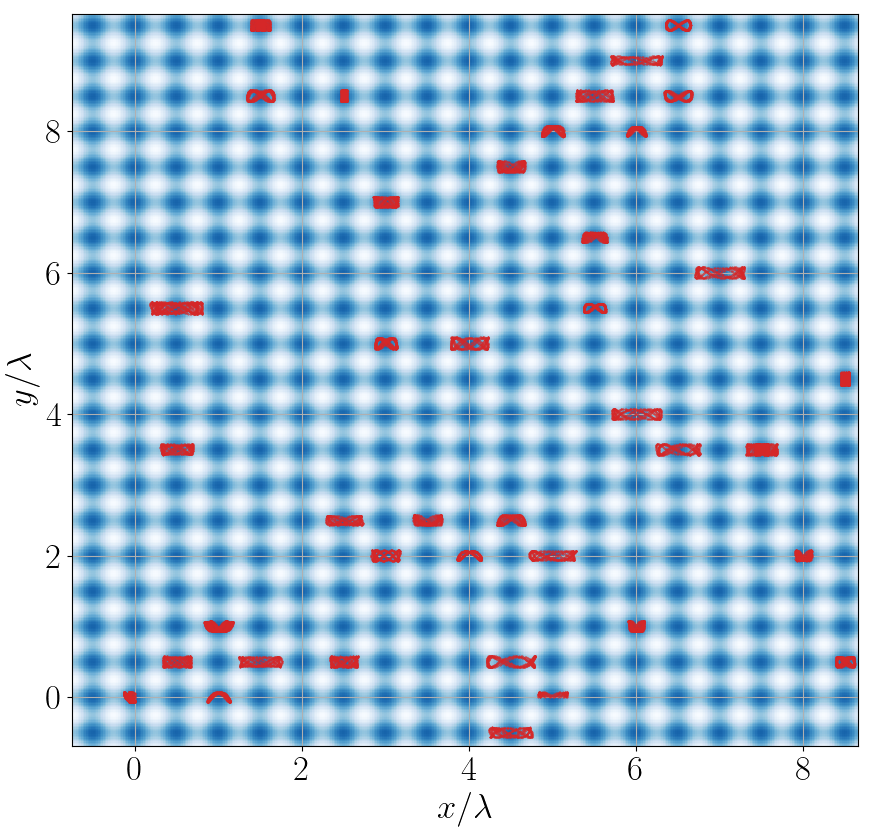}
	\caption{\textit{Self-Organization.} Atomic trajectories in the $xy$-plane (red color) during the final $30\Gamma$ time period of the dynamics in the regime of self-organization presented in fig.~\ref{fig1}. The trajectories are plotted on top of the light intensity distribution to show the emerging checkerboard pattern of trapped atoms in the $xy$-plane.}
	\label{fig1a}
\end{figure}

The full master equation description does not allow to numerically calculate many atoms due to the exponential growth of the Hilbert space with the atom number. Hence, in order to simulate the system dynamics and spectrum of the light field in the cavity for an ensemble of many atoms we use a cumulant expansion approach \cite{kubo62, Holland09, plankensteiner2022quantumcumulants}. We restrict ourselves to the second-order cumulant expansion and assume that the higher-order correlations are negligible, i.~e. $\langle A B C \rangle_C \approx \langle A \rangle \langle B C \rangle +\langle B \rangle \langle A C \rangle + \langle C \rangle \langle A B \rangle - 2 \langle A \rangle \langle B \rangle \langle C \rangle$. Thus, we start from the Heisenberg equations for operator averages describing our system, which for a given operator $\cal{O}$ reads
\begin{equation}
\label{ceq_O}
\frac{d}{dt} \langle {\mathcal{O}} \rangle = i \langle [H, {\cal{O}}] \rangle + \kappa\langle\mathcal{D}[a]\mathcal{O}\rangle +\Gamma\sum_i\langle\mathcal{D}[\sigma_i^-]\mathcal{O}\rangle,
\end{equation} 
where $\mathcal{D}[c]\mathcal{O} = \left(2c^\dagger\mathcal{O}c - c^\dagger c\mathcal{O} - \mathcal{O}c^\dagger c\right)/2$. Using the second-order cumulant expansion we obtain a closed set of equations describing our system,
\begin{widetext}
\begin{equation}
\label{eq.Heisenberg}
\begin{aligned}
&\frac{d}{dt} \langle a \rangle =  -(\kappa/2-i\Delta_c) \langle a \rangle -i\sum_j g(x_j,y_j) \langle \sigma^-_j \rangle \cr
&\frac{d}{dt} \langle \sigma^-_m \rangle = -(\Gamma/2-i\Delta_{a})\langle \sigma^-_m \rangle +ig(x_m,y_m) \langle a \rangle (2\langle \sigma^+_m \sigma^-_m \rangle -1) +i\Omega \cos(k y_m)(2\langle \sigma^+_m \sigma^-_m \rangle-1)\\
&\frac{d}{dt} \langle a^\dagger a \rangle = -\kappa \langle a^\dagger a \rangle + i\sum_j g(x_j,y_j)  (\langle a \sigma^+_j \rangle - \langle a^\dagger \sigma^-_j \rangle) \\
&\frac{d}{dt} \langle a \sigma^+_m \rangle = -(\frac{\kappa +\Gamma}{2}+i\Delta_{a}-i\Delta_c)\langle a \sigma^+_m \rangle + ig(x_m,y_m)(\langle a^\dagger a \rangle-2\langle a^\dagger a \rangle \langle \sigma^+_m \sigma^-_m \rangle-\langle \sigma^+_m \sigma^-_m \rangle) \\
&\quad \quad \quad -i \sum_{j;m\neq j} g(x_j,y_j)\langle \sigma^+_m \sigma^-_j \rangle -i\Omega \cos(k y_m)\langle a \rangle(2\langle \sigma^+_m \sigma^-_m \rangle-1) \\
&\frac{d}{dt}  \langle \sigma^+_m \sigma^-_m \rangle = -\Gamma \langle \sigma^+_m \sigma^-_m \rangle -ig(x_m,y_m)(\langle a \sigma^+_m \rangle - \langle a^\dagger \sigma^-_m \rangle)
-i\Omega\cos(k y_m)(\langle \sigma^+_m \rangle-\langle \sigma^-_m \rangle) \\
&\frac{d}{dt}  \langle \sigma^+_m \sigma^-_j \rangle = -\Gamma \langle \sigma^+_m \sigma^-_j \rangle -ig(x_m,y_m)\langle a^\dagger \sigma^-_j \rangle (2\langle \sigma^+_m \sigma^-_m \rangle -1) +ig(x_j,y_j)\langle a \sigma^+_m \rangle (2\langle \sigma^+_j \sigma^-_j \rangle -1)\\
&\quad \quad \quad +i\Omega \cos(k y_j)\langle \sigma^+_m \rangle (2 \langle \sigma^+_j \sigma^-_j \rangle -1) -i\Omega \cos(k y_m)\langle \sigma^-_j \rangle (2 \langle \sigma^+_m \sigma^-_m \rangle -1) \\
&\frac{d}{dt}  \langle x_m \rangle = 2\omega_r \langle p_{xm} \rangle / k^2  \\
&\frac{d}{dt}  \langle y_m \rangle = 2\omega_r \langle p_{ym} \rangle / k^2  \\
&\frac{d}{dt}  \langle p_{xm} \rangle = 2g k \sin(k x_m) e^{-y_m^2/w_0^2} \Re \{\langle a \sigma^+_m \rangle \}\\
&\frac{d}{dt}  \langle p_{ym} \rangle = \frac{4g}{w_0^2}\cos(k x_m) y_m e^{-y_m^2/w_0^2} \Re \{\langle a \sigma^+_m \rangle \} +2k \Omega\sin(k y_m) \Re \{\langle \sigma^+_m \rangle \},
\end{aligned}
\end{equation}
\end{widetext}
where $m=1..N$, $g(x_m, y_m) = g \cos{(k\langle x_m \rangle)} e^{- \langle y_m\rangle ^2/w_0^2}$, and $\Re \{ \langle \cal{O} \rangle \}$ is used to denote the real part of an expectation value of an operator. 
Here we neglect the effects of momentum diffusion associated with spontaneous photon emission, which generally may play a role in the self-organization dynamics. However, the relevance of these effects on the final temperature and self-organization threshold strongly depends on the operating parameter regime \cite{asboth2005self, niedenzu2011kinetic}. In particular, we assume the pump laser detuning from the atomic transition frequency to be larger than the pumping strength $\Delta_a > \Omega > \Gamma$ to avoid the heating from a strong laser drive.  Throughout the considered regime the excited state population might not be negligible but still remains rather small. 
Moreover, in a clock atom ensemble the spontaneous emission rate is typically much smaller than the other rates appearing in the model, ultimately being many orders of magnitude smaller than the cavity decay rate. In this case, the occurrence of spontaneous emission events in the system is already very rare on the time scales of the atomic dynamics, such that we expect only a minor contribution to the dynamics.
Starting from the initial atomic positions placed at the center of the cavity, we neglect direct interatomic interactions and assume that dipole-dipole interactions and collisions only become significant for dense atomic ensembles~\cite{Hotter19}. The initial atomic momenta are randomly distributed in the range of several $\hbar k$. The resulting solution gives us the time evolution of the atomic positions and momenta, as well as the mean intracavity photon number $\langle a^{\dagger}a \rangle$ and the population inversion averaged over the atomic ensemble $\overline{\langle s_z \rangle} = \frac{1}{N}\sum_i \langle s_z^i \rangle$, where $\langle \sigma_z^i \rangle = \langle \sigma^+_i \sigma^-_i \rangle - \langle \sigma^-_i \sigma^+_i \rangle$, which are shown in fig.~\ref{fig1} for $N=100$ atoms. In the final stage of the time evolution, when the system dynamics has stabilized, one can plot the atomic trajectories in the $xy$-plane to demonstrate the emerging checkerboard pattern in the self-organization regime, as presented in fig.~\ref{fig1a}.

\begin{figure*}[ht!]
    \centering
    	\includegraphics[width=\columnwidth]{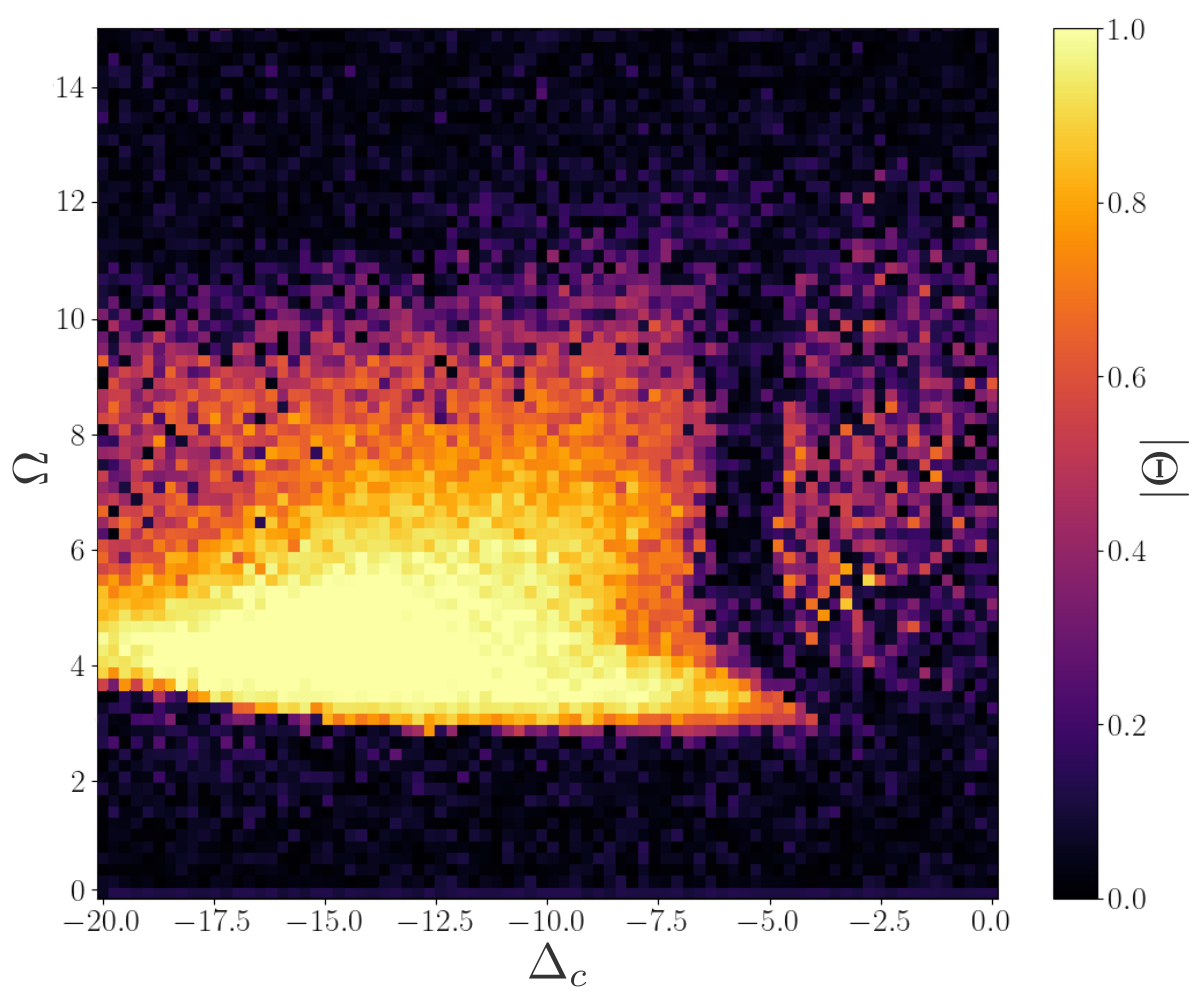}
    \includegraphics[width=\columnwidth]{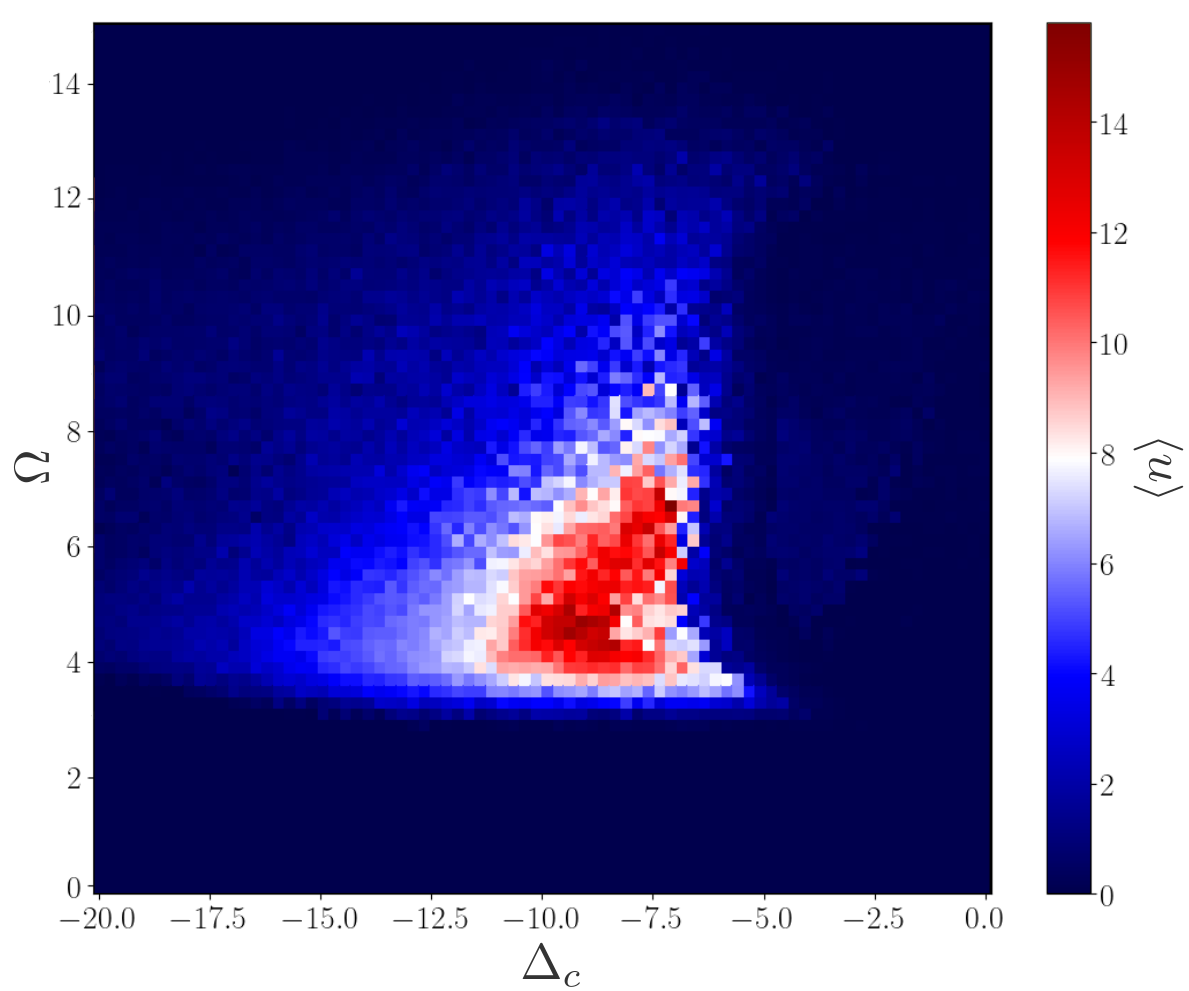}
	\caption{\textit{Self-Organization (Laser Drive).} Absolute value of the order parameter (left) and mean photon number (right) scans for $N=100$ atoms depending on the cavity detuning ($\Delta_c$) and driving strength ($\Omega$). Parameters: $\kappa = 10\Gamma$, $g = 1.5\Gamma$, $\Delta_a = -20\Gamma$, $w_0 = 1000 \lambda$, $\omega_r = 1\Gamma$.}
	\label{fig2}
\end{figure*}

\begin{figure*}[ht!]
    \centering
	\includegraphics[width=\columnwidth]{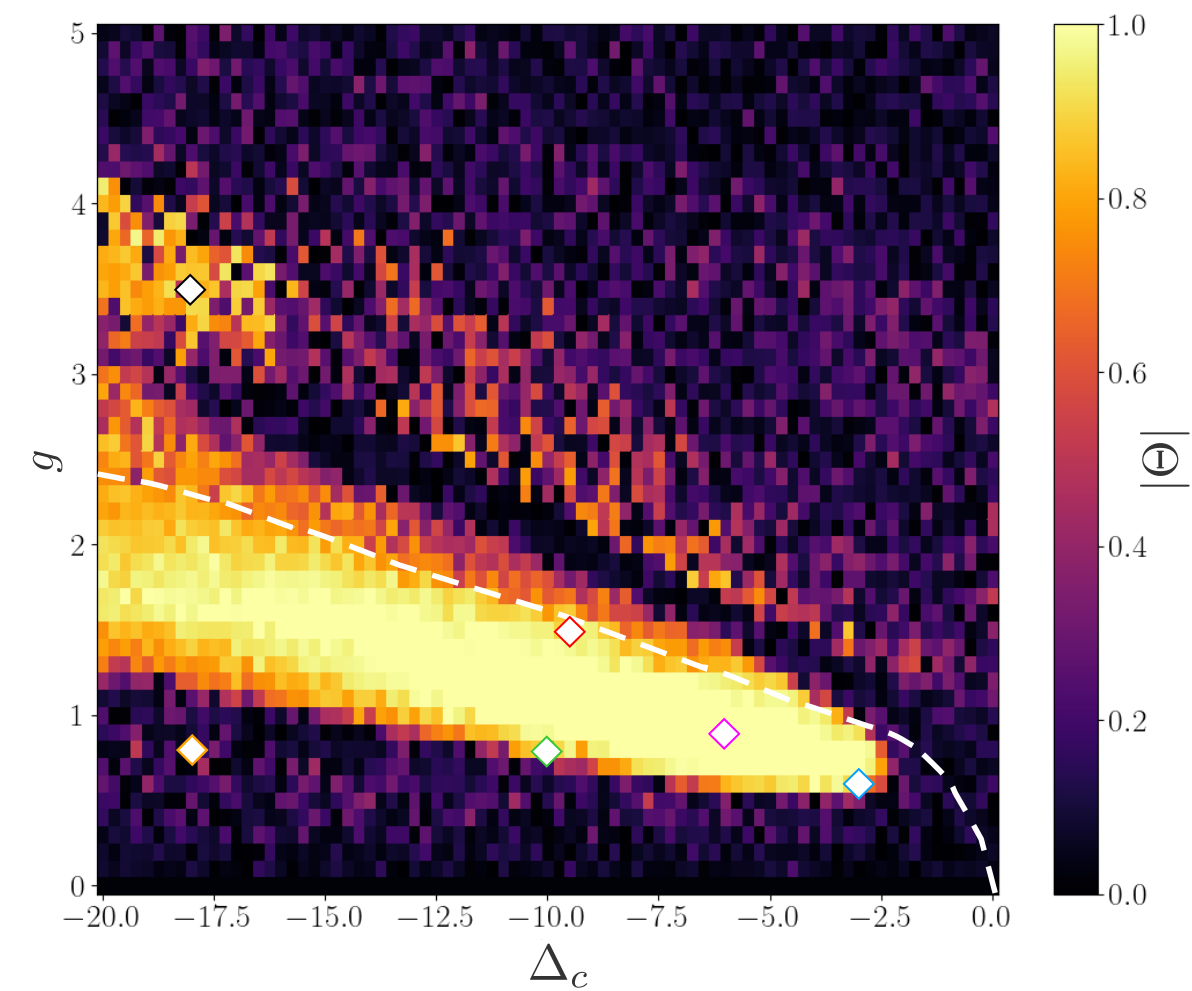}
    \includegraphics[width=\columnwidth]{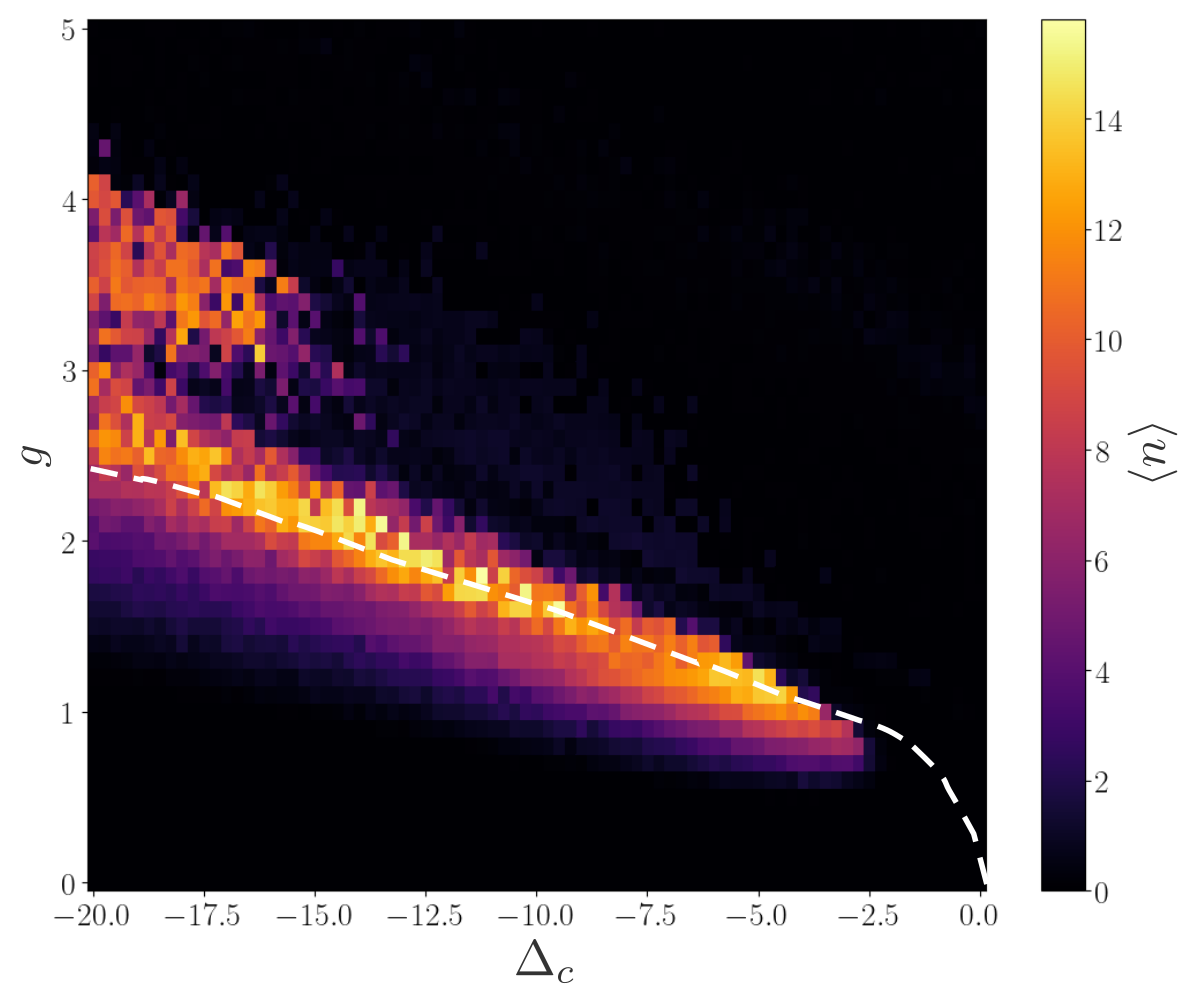}
	\caption{\textit{Self-Organization (Cavity Coupling).} Absolute value of the order parameter (left) and mean photon number (right) for $N=100$ atoms depending on the cavity detuning ($\Delta_c$) and atom-cavity coupling strength ($g$). Parameters: $\kappa = 10\Gamma$, $\Omega = 5\Gamma$, $\Delta_a = -20\Gamma$, $w_0 = 1000 \lambda$, $\omega_r = 1\Gamma$. The white dashed line indicates the zeros of the effective cavity detuning calculated at each point according to eq.~(\ref{condition_1}). White rhombus markers show different parameter sets used in fig.~\ref{fig3-4} for the calculation of the cavity field spectra.}
	\label{fig3}
\end{figure*}

In fig.~\ref{fig2} we calculate the time-averaged order parameter and mean photon number in the cavity mode during the final stage of the dynamics as a function of the cavity detuning and the driving strength. In order to obtain a comprehensive parameter scan we restrict ourselves to the first-order cumulant expansion (mean-field solution), which does not allow for calculating the spectral properties of the light field but substantially reduces the number of equations in eqs.~(\ref{eq.Heisenberg}). Both the second-order and the mean-field solution are in good agreement with each other as well as with the full master equation solution for small atom numbers. Above the pumping threshold one can observe self-ordering, where particles organize in a regular checkerboard pattern characterized by the order parameter
\begin{equation}
\label{order_parameter}
\Theta = \frac{1}{N}\sum_{i=1}^N \cos (\langle k x_i \rangle) \cos (\langle k y_i \rangle) \longrightarrow \pm 1
\end{equation} 
corresponding to one of the two possible realizations of the checkerboard. Increasing the driving strength further towards saturation leads to cavity heating and the expulsion of atoms from the cavity. Therefore, in the following we choose an intermediate optimal value of the driving Rabi frequency and present the scan over the atom-cavity coupling strength for $\Omega = 5\Gamma$ in fig.~\ref{fig3}. The white dashed line indicates the zeros of the effective cavity detuning in eq.~(\ref{condition_1}) calculated at each point of the parameter scan. Across the line, the effective cavity detuning changes its sign and becomes positive resulting in cavity heating. White rhombus markers indicate different parameter choices used for the calculation of the cavity field spectra shown in fig.~\ref{fig3-4}, which is discussed in the following section.

\begin{figure}[ht!]
	\includegraphics[width=\columnwidth]{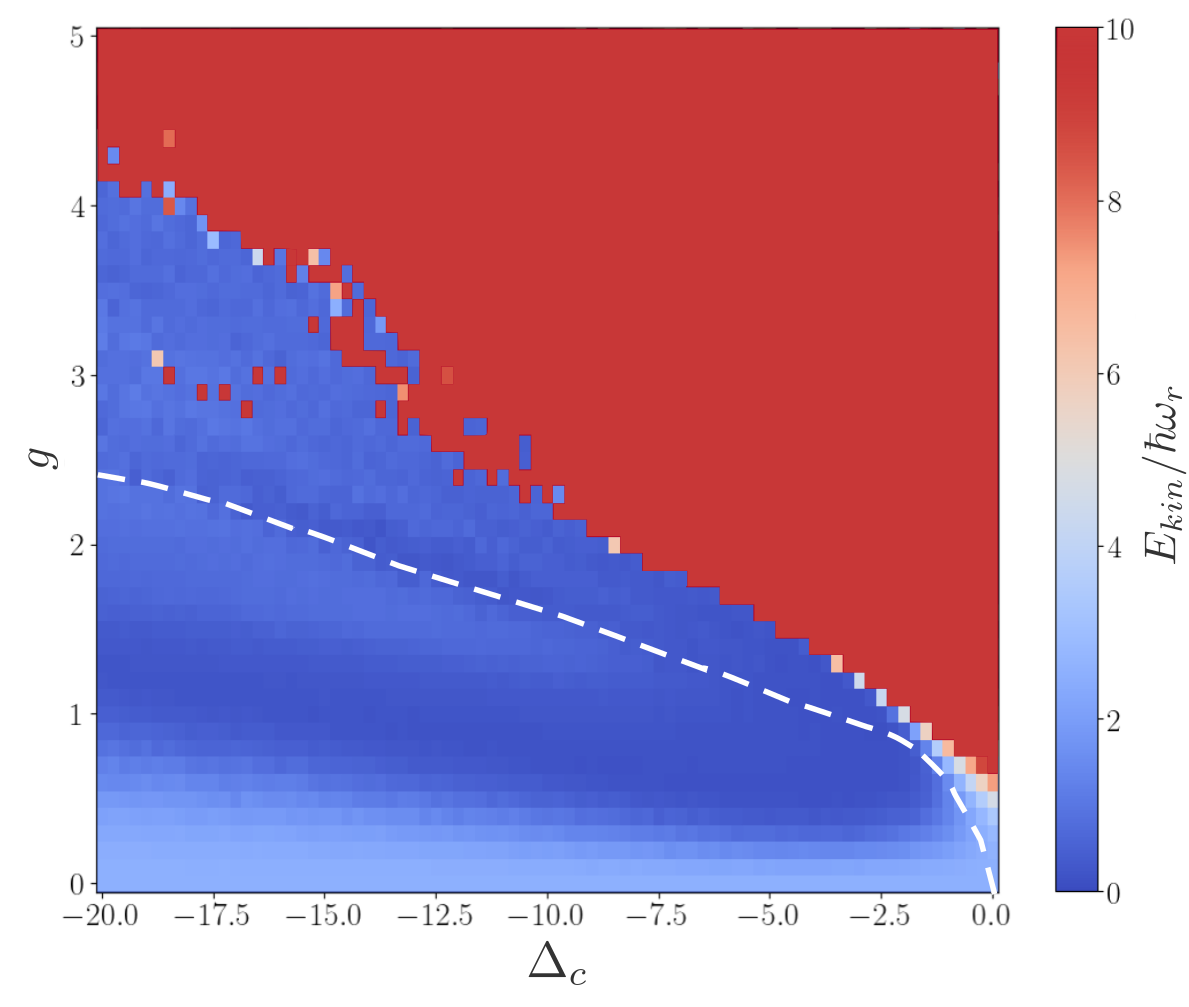}
	\caption{\textit{Cavity Cooling.} Kinetic energy in eq.~(\ref{E_kin}) presented as a function of the cavity detuning ($\Delta_c$) and atom-cavity coupling strength ($g$). The kinetic energy upper limit is set to $10 \hbar \omega_r$ for a better resolution in the cooling region. The parameters are the same as in fig.~\ref{fig3} with the white dashed line indicating the zeros of the effective cavity detuning.}
	\label{fig_E_Kin}
\end{figure}

In fig.~\ref{fig_E_Kin} we present the kinetic energy per atom,
\begin{equation}
\label{E_kin}
\langle E_{kin} \rangle = \frac{1}{N} \sum_{i=1}^N \frac{\langle p_{xi} \rangle^2 + \langle p_{yi} \rangle^2}{2m},
\end{equation} 
which expectation value is time-averaged during during the final stage of the atomic dynamics. One can clearly see the sharp transition between the atomic cooling and heating regimes. However, this transition happens significantly further away than one would expect from the zeros of the effective cavity detuning indicated by the white dashed line. In the upper left region of the plot, where the cavity frequency approaches the atomic frequency, one can observe an extended region of cavity cooling and self-organization. Here, the cavity field as well as the atomic dynamics become more noisy but ordering persists.

%
%
%

%
%
%
%

%
%

\section{Cavity output spectrum}

%
\begin{figure}[t!]
    \centering
	\includegraphics[width=\columnwidth]{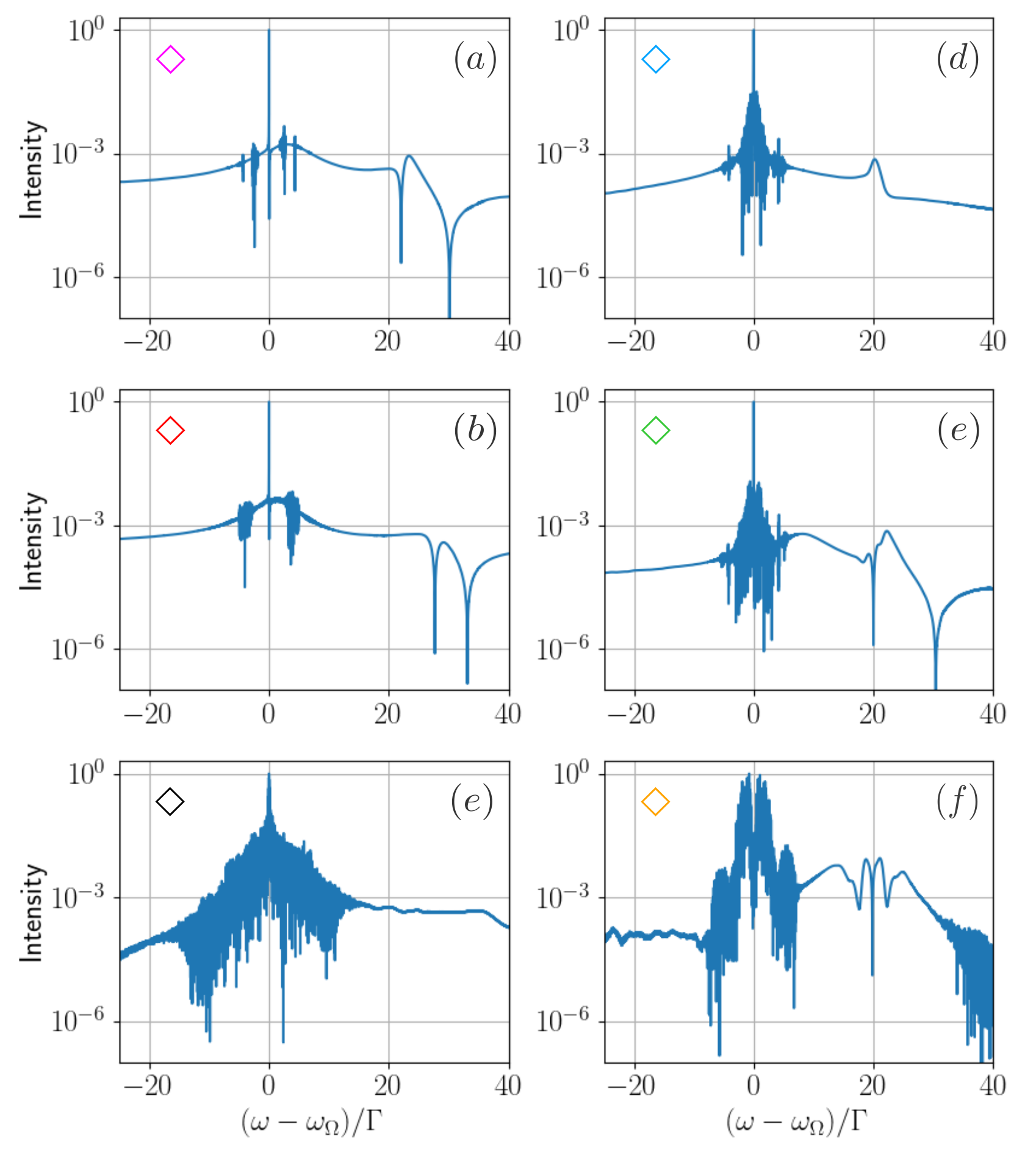}
	\caption{\textit{Spectra.} Normalized spectral distribution of the cavity output field for $N=100$ atoms in different parameter regions depicted by white rhombus markers in fig.~\ref{fig3}. An increased spectral intensity and a very narrow dip in the spectrum appear in the vicinity of the bare atomic transition frequency close to the self-organization threshold. Parameters: (a) $\Delta_c = -6\Gamma$, $g=0.9\Gamma$; (b) $\Delta_c = -9.5\Gamma$, $g=1.5\Gamma$; (c) $\Delta_c = -18\Gamma$, $g=3.5\Gamma$; (d) $\Delta_c = -3\Gamma$, $g=0.6\Gamma$; (e) $\Delta_c = -10\Gamma$, $g=0.8\Gamma$; (f) $\Delta_c = -18\Gamma$, $g=0.8\Gamma$, $\kappa = 10\Gamma$, $\Omega = 5\Gamma$, $\Delta_a = -20\Gamma$, $w_0 = 1000 \lambda$, $\omega_r = 1\Gamma$.}
	\label{fig3-4}
\end{figure}
%

%
%

Even in the regime when the system dynamics has stabilized to a quasi-stationary state, the atomic motion creates additional fluctuations of the photon number around its average value. These fluctuations are significant for a small number of particles in the simulation but become less pronounced the larger the atom number is.  In order to find the spectrum we make use of the quantum regression theorem~\cite{Carmichael13} and calculate the first-order correlation function $g^{(1)}(\tau) = \langle a^{\dagger}(t_0+\tau)a(t_0) \rangle$. According to the Wiener-Khinchin theorem \cite{Puri01} the spectrum can be found as the Fourier transform of the first-order correlation function,
\begin{equation}
\label{eq_spectrum}
S(\omega) = 2 \Re \left\{\int_{0}^{\infty} d\tau e^{-i\omega \tau} g^{(1)} (\tau) \right\}.
\end{equation} 

We examine the cavity output spectrum at each point of the scan in fig.~\ref{fig3} searching for narrow emission or lasing close to the bare atomic transition frequency with a particular interest in the region around self-organization. We use the second-order cumulant approach in eqs.~(\ref{eq.Heisenberg}) combined with the equations for the dynamics of the first-order correlation function in order to compute the spectrum~\cite{Bychek2021}. In fig.~\ref{fig3-4} we present the spectra of the cavity light field for an ensemble of $N=100$ atoms obtained for different parameter sets indicated by white rhombus markers in fig.~\ref{fig3}.

In the ordered regime, a coherent peak at the pump frequency is dominant in the spectra, accompanied by motional broadening. In the case of perfect ordering presented in the spectrum in fig.~\ref{fig3-4}(a), one can resolve the  motional sideband peaks as all atoms in the ensemble oscillate around their trapping positions in the same way. Additionally, in fig.~\ref{fig3-4}(b) we present the spectrum in the area of the maximal photon number given in fig.~\ref{fig2} for $g=1.5\Gamma$. Above the main region of the self-organization, both the cavity field and atomic dynamics become rather noisy. Here, we do not observe any spectral features close to the atomic frequency even in the extended region of self-organization, as shown in fig.~\ref{fig3-4}(c).

Of particular interest are the spectra in the parameter region where the self-ordering threshold has not been fully reached. In fig.~\ref{fig3-4}(d)-(f) we observe an increased spectral intensity close to the atomic transition frequency, which we associate with the emission from the excited atomic fraction. Because the atoms in this region are not perfectly ordered, they undergo complex dynamics as they continue to move, oscillate, and jump from one field antinode to another. Note, that the dipole light force has the opposite sign depending on the atomic state. Thus, in the case of the red-detuned laser drive, the ground-state atoms are drawn to the field maxima, whereas the excited-state atoms are expelled from these regions. In this case, the excited atoms can subsequently move through the field minima and emit photons that will be visible in the spectrum near the bare atomic frequency. Moreover, in fig.~\ref{fig3-4}(e)-(f), we find that the spectra reveal a narrow absorption minimum at the atomic frequency, which can be associated with the atomic antiresonance~\cite{alsing1992suppression, sames2014antiresonance}. 
In the self-organization regime, the position of this minimum becomes shifted above the bare atomic frequency. We attribute this to the energy light shifts that atoms experience being trapped around the field antinodes.

%
\begin{figure}[h!]
    \centering
	\includegraphics[width=\columnwidth]{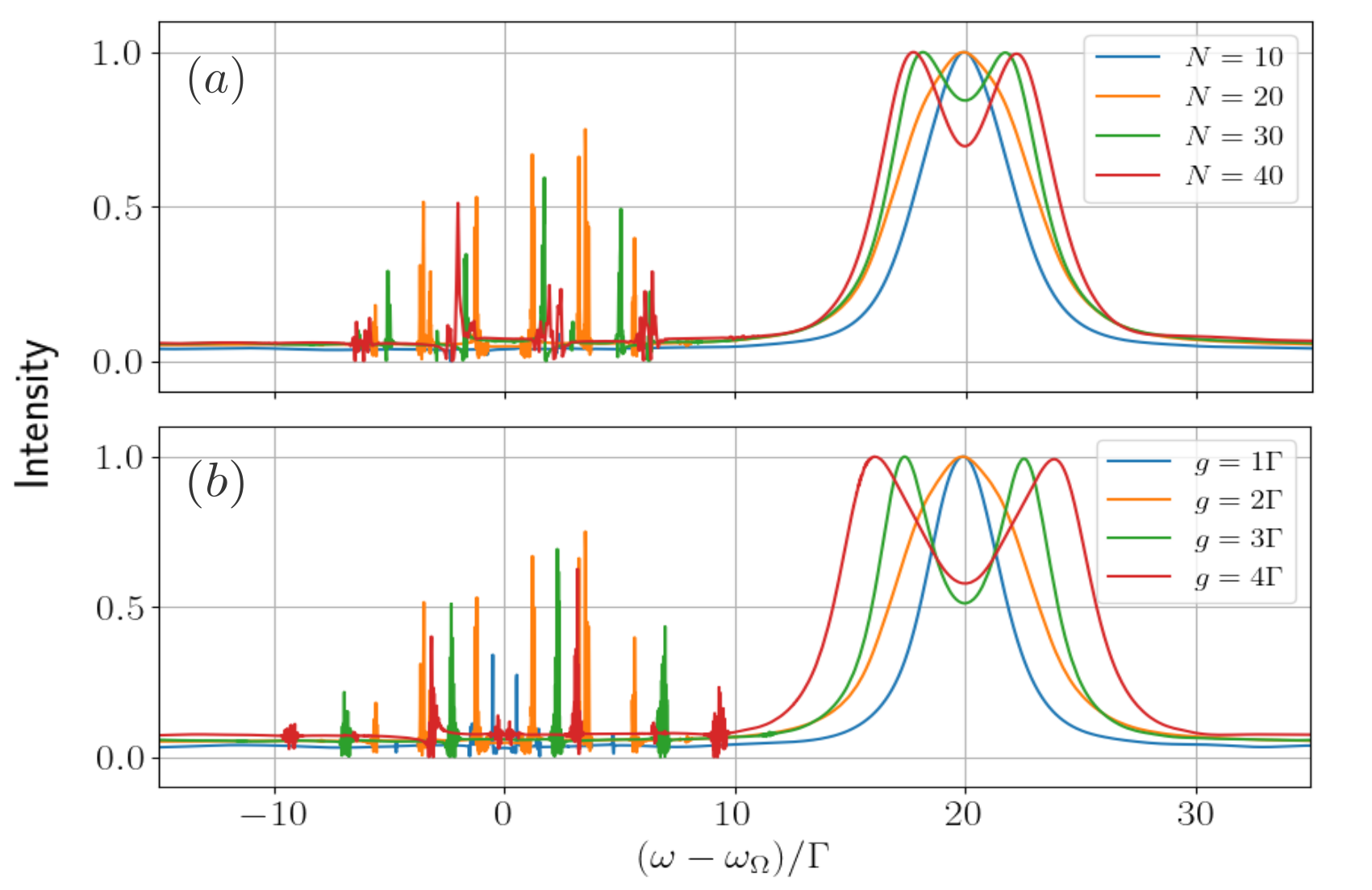}
	\caption{\textit{Filter Cavity Mode Spectra.} (a) Spectra of the filter cavity mode output for a coupling strength $g=2\Gamma$ and different atom numbers $N$. (b) Same as above for varying coupling strength $g$ and $N=20$ atoms. The operating parameters are: $\kappa = 10\Gamma$, $\Omega = 5\Gamma$, $\Delta_a = \Delta_{c2} = -20\Gamma$, $\Delta_c = -10\Gamma$, $w_0 = 1000 \lambda$, $\omega_r = 1\Gamma$. }
	\label{fig4}
\end{figure}
%

Along with the main cosine mode of the cavity it is interesting to study the spectra when we introduce an additional sine mode into the system resonant to the atomic transition frequency. This sine mode will play the role of a filter for the light coming from the atoms. Interesting behavior occurs when the atoms are close to but not quite in the self-organization regime as they not only oscillate around their position as in fig.~\ref{fig1a} but can jump from one antinode of the field to another, thereby increasing the emission probability into the sine mode. Of course, there is much less light going into the filter mode than there is light in the main mode of the cavity.

The second mode makes the simulation for the spectra much more computationally demanding substantially increasing the amount of equations in the system described by the Hamiltonian
\begin{equation}
\begin{aligned}
    \label{eq:H}
    H' = H-\Delta_{c2} b^{\dagger}b + \sum_{i=1}^N g \sin(k x_i) e^{-y_i^2/w_0^2} (b \sigma^+_i + b^\dagger \sigma^-_i),
\end{aligned}
\end{equation}
where $\Delta_{c2} = \omega_{\Omega}-\omega_{c2} $ and $b^\dagger$, $b$ are the photon creation and annihilation operators in the second cavity mode.
In fig.~\ref{fig4}(a) we fix the cavity coupling strength to $g=2\Gamma$ and increase the atom number up to $N=40$ atoms. One can observe the normal mode splitting when the condition $g\sqrt{N}>\kappa$ is fulfilled. Similarly, fixing the number of atoms and changing the coupling constant leads to a splitting in spectra depicted in fig.~\ref{fig4}(b).

\section{Conclusions}

In this work, we have studied spatial self-organization and spectral properties of the emitted light in a transversely driven clock atom ensemble coupled to a linear Fabry-Perot optical cavity. Our simulations demonstrate that cavity cooling and atomic self-ordering via collective scattering also appear in the bad-cavity limit where the atomic linewidth is much smaller than the cavity linewidth and of the order of the recoil frequency. Under a strong drive, state-dependent light forces induced by the driving laser and the light scattered into the cavity allow for the atoms to acquire excited state population while subsequently moving through the field's minima. In the self-organization regime, a dominant spectral peak in the cavity output stems from coherent light scattering at the pump frequency and is surrounded by motional sidebands. When the self-organization threshold has not been fully reached, we observe an increased spectral intensity as well as a narrow antiresonance close to the unperturbed atomic transition frequency, which can be useful for clock spectroscopy.
Finally, in order to provide for a continuous frequency reference, the light coming from the excited atomic fraction can be filtered via a second cavity mode spatially shifted by a quarter wavelength and tuned to the bare atomic transition frequency. We anticipate that this will lead to laser-like emission for sufficiently large ensembles, which, unfortunately, goes beyond our simulation capabilities, but is readily available in experiment~\cite{schafer2024continuous}.

\section*{Data availability}

Numerical simulations were performed with the open-source framework Differentialequations.jl~\cite{rackauckas2017differentialequations} and QuantumOptics.jl~\cite{kramer2018quantumoptics} in the Julia programming language. The graphs were produced using the Matplotlib library~\cite{hunter2007matplotlib}. All raw data are openly available at \href{https://doi.org/10.5281/zenodo.12796758}{\it{https://doi.org/10.5281/zenodo.12796758}}.\\

\section*{Acknowledgments}

We would like to thank Vera Sch{\"a}fer for helpful discussions and her comments on the manuscript. We acknowledge funding from the L'ORÉAL Austria Fellowship "For Women in Science" 2023 (A.B.) and the FET OPEN Network Cryst3 funded by the European Union (EU) via Horizon 2020 (L.O., H.R.). This work was also financed with funds from the state of Tyrol (A.B., L.O.).\\

\bibliography{Superradiant_laser}

\end{document}